\documentclass[prb,superscriptaddress,twocolumn]{revtex4-2}
\usepackage{graphicx}
\usepackage{amsmath}
\usepackage{braket}
\usepackage{color}
\usepackage{float}
\begin{document}

\title{ LaVO$_3$: a true 
Kugel-Khomskii system}
\author{Xue-Jing Zhang}
\affiliation{
Institute for Advanced Simulation, Forschungszentrum J\"ulich, 
52425 J\"ulich, Germany}
\author{Erik Koch}
\affiliation{
Institute for Advanced Simulation, Forschungszentrum J\"ulich, 
52425 J\"ulich, Germany}
\affiliation{
JARA High-Performance Computing, 52062, Aachen, Germany.}
\author{Eva Pavarini}
\affiliation{
Institute for Advanced Simulation, Forschungszentrum J\"ulich, 
52425 J\"ulich, Germany}
\affiliation{
JARA High-Performance Computing, 52062, Aachen, Germany.}
\date{\today}%

\begin{abstract}
We show that  the $t_{2g}^2$ perovskite LaVO$_3$, in its orthorhombic phase, is a rare case 
of a system %
{hosting an  orbital-ordering Kugel-Khomskii phase transition,
rather than being controlled by the Coulomb-enhanced crystal-field splitting.}
We find  that, as a consequence of this, the magnetic transition is close to (and even above) the super-exchange driven orbital-ordering transition, whereas typically
magnetism arises at much lower temperatures than orbital ordering.
Our results support the experimental scenario of orbital-ordering and G-type spin correlations just above the monoclinic-to-orthorhombic  structural change.
To explore the effects  of crystal-field splitting and filling, we compare to YVO$_3$ and $t_{2g}^1$ titanates.
{In all these materials the crystal-field is sufficiently large to suppress the Kugel-Khomskii phase transition.}
 \end{abstract}

\maketitle

\section{Introduction}
Almost 50 years ago, Kugel and Khomskii showed in a classic paper that, in strongly-correlated materials,
orbital ordering (OO) can arise from pure super-exchange (SE) interactions \cite{KK}. 
It can, however, also result from the crystal-field (CF)  splitting via a lattice distortion, i.e., from  electron-lattice coupling \cite{Kanamori}. In typical cases both mechanisms lead to similar types of ordering, so that identifying which one actually drives the transition is a ``chicken-and-egg problem'' \cite{DK}.
Despite the intensive search, it has therefore been hard to find an undisputed realization of a Kugel-Khomskii  (KK) system.

Initially it  was believed that 
the $e_g$ perovskites KCuF$_3$ and LaMnO$_3$ could be KK materials \cite{DK,K,La,Faz}.
In recent years, however, it was proven that neither in these nor in other $e_g$ systems, super-exchange 
interactions are strong enough to drive the OO transition alone \cite{prl2008,prl2010,autieri,lamno3c,tote,julian1}.
In fact, in order to explain the presence of OO at high temperature,
lattice distortions, arising from the Jahn-Teller effect, the Born-Meyer
potential or both \cite{Hunter}, have to be present. The CF splitting generated by the distortions is then effectively enhanced
by the Coulomb repulsion, which suppresses orbital fluctuations \cite{3d1a,3d1b,Molly2007}, leading
to a very robust OO state. The actual form of the ordered state is then essentially determined by the crystal field.
\begin{figure}[t]
\centering  
{\hspace*{-0.25cm}\includegraphics[width=0.5\textwidth]{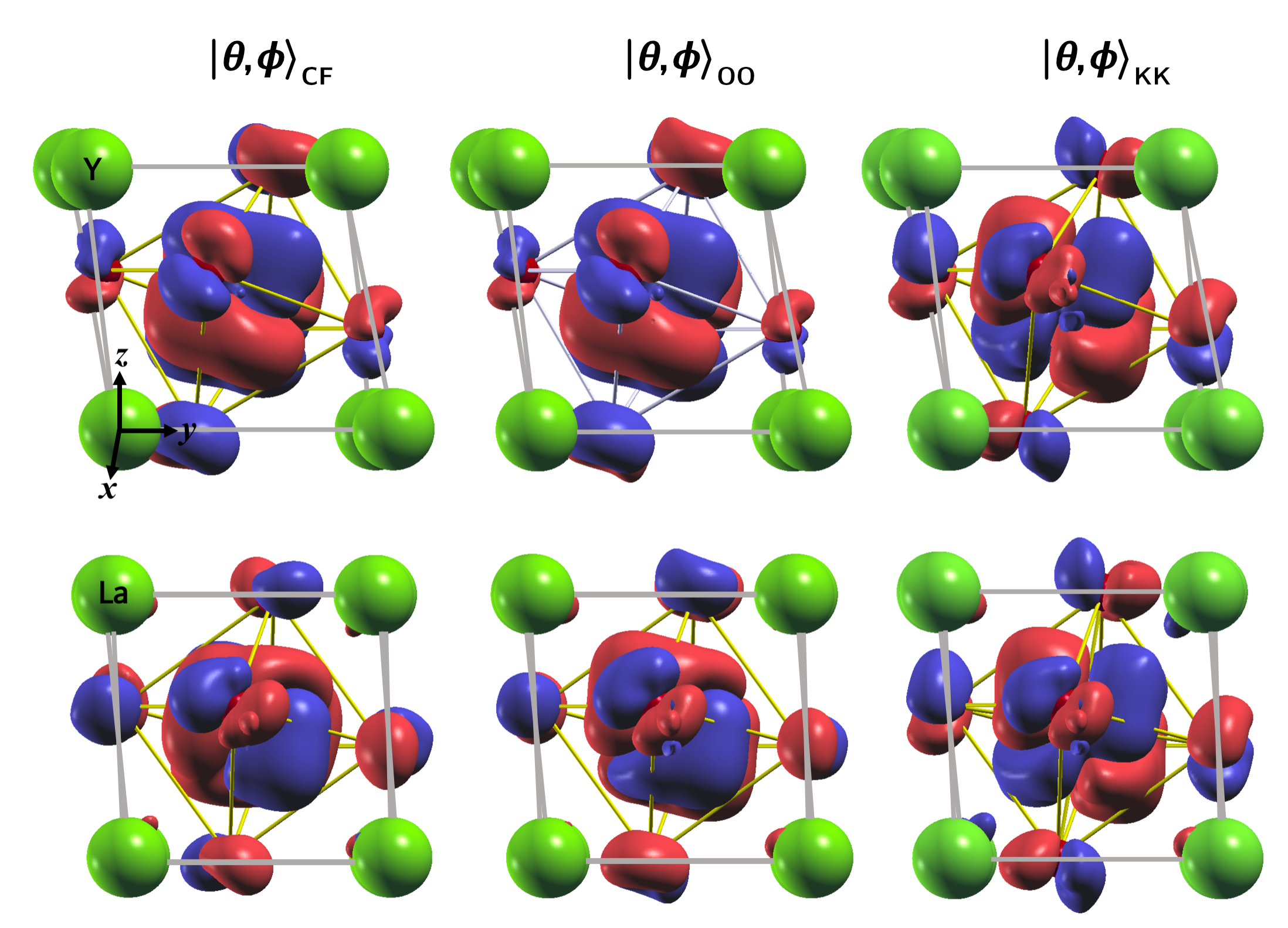}}
\caption{ \label{orbs}
 Left: LDA,  highest energy crystal-field (CF) state  $|\theta,\phi\rangle_{\rm CF}$.  
 Center and right: DMFT  hole orbital at 90~K.
$|\theta,\phi\rangle_{\rm OO}$:  experimental structure.
$|\theta,\phi\rangle_{\rm KK}$:   idealized case with no CF splitting. Top line: YVO$_3$. Bottom line: LaVO$_3$.  
}
\end{figure}
Recently, the case of $t_{2g}^1$ perovskites was explored as possible alternative  \cite{xj2020}. In fact, 
the CF  is  typically weaker
in $\pi$-bond   than in $\sigma$-bond materials, while, at the same time,
the orbital degeneracy is larger. This can  potentially turn the balance of interactions
 in favor of SE. It was indeed shown that $T_{\rm KK}$,
the critical temperature of super-exchange driven orbital ordering, is remarkably large
in LaTiO$_3$ and YTiO$_3$, about 300 K \cite{xj2020}.
At the same time it became, however, also clear  that even in these systems OO is
 dominated by correlation-enhanced CF splitting rather than
 the SE interaction.

In this paper we identify the first compound in which KK multi-orbital super-exchange
{{yields an orbital-ordering phase transition at observable temperatures}}:
the $t_{2g}^2$ perovskite LaVO$_3$ in its orthorhombic (GdFeO$_3$-type structure) phase.
LaVO$_3$ has drawn a lot of attention due to its peculiar properties and phase diagram 
\cite{Bordet1993,Miyasaka2003,Yan2004,Yan2011,Ren2003,Zhou2008,Tung2008}.  
In the low-temperature  monoclinic phase $(T<T_{\rm str} \sim 140$~K), it exhibits a C-type magnetic structure.
Due to the small distortions of the VO$_6$ octahedra, it was suggested that 
C-type spin order  could arise from strong $xz/yz$ orbital fluctuations \cite{Khal}.
Calculations accounting for  the GdFeO$_3$-type distortions have  however shown that orbital fluctuations are suppressed below $T_{\rm str}$,
and classical OO is sufficient to explain C-type spin order \cite{Molly2007}.
The debate remains open for what concerns the intriguing phase just above the structural phase transition,
 $T_{\rm str}< T< T_{\rm N^\prime}{\sim} 145$~K. 
Here thermodynamic anomalies and weak magnetic peaks
suggest a change in spin-orbital  order, either long or short ranged \cite{Ren2003,Zhou2008,Tung2008},
the origin of which is, however, unclear. 
Our results show that  the KK nature of LaVO$_3$ provides a natural explanation.

To understand what makes  orthorhombic LaVO$_3$  special, we compare it with the similar but more distorted YVO$_3$ and  the two iso-structural $t_{2g}^1$   titanates.
We start from the paramagnetic  (PM) phase. We show that, for the vanadates, $T_{\rm KK}$ 
is actually smaller than for titanates, $T_{\rm KK}\sim 190$~K 
in LaVO$_3$ and $T_{\rm KK}\sim 150$~K in YVO$_3$, but
at the same time, the crystal field  is weaker, for LaVO$_3$ more so than for YVO$_3$.
The surprising result of the energy-scale balance is that, while in YVO$_3$ the occupied state 
is still, to a large extent, determined by the CF splitting,  for LaVO$_3$ it substantially differs from CF-based predictions (Fig.~\ref{orbs}).
Staying with LaVO$_3$,  we find that the actual orbital-ordering temperature, $T_{\rm OO}$, is close to $T_{\rm KK}$.
Even more remarkable is the outcome of magnetic calculations. They yield a G-type  antiferromagnetic (AF)  phase
with $T_N> T_{\rm OO}\sim T_{\rm KK}>T_{\rm str}$. 
This is opposite to what typically happens  in  orbitally-ordered materials, %
in which $T_N$ is smaller
 than $T_{\rm OO}$, often sizably so. Our results provide a microscopic explanation of the spin-orbital correlations found right above $T_{\rm str}$ in experiments  \cite{Ren2003,Zhou2008,Tung2008}.

\section{Orbital-ordering transition}
In order to determine the onset of the super-exchange driven orbital-ordering transition, $T_{\rm KK}$, we adopt the approach pioneered in Refs.~\cite{prl2008,prl2010}.
It consist in progressively reducing the static CF splitting to single out the effects of pure super-exchange 
from those of Coulomb-enhanced structural distortions.
Calculations are performed using the local-density approximation plus dynamical-mean-field theory method (LDA+DMFT) \cite{ldadmft}
for the materials-specific $t_{2g}$ Hubbard model
\begin{align} \nonumber  \label{Hub}
\hat{H}=&-\sum_{ii'\sigma} \sum_{mm'} t^{i,i'}_{mm'}c^{\dagger}_{im\sigma}c_{i'm'\sigma}+U\sum_{im}\hat{n}_{im\uparrow}\hat{n}_{im\downarrow} \\ 
&+\frac{1}{2}\sum_{i\sigma \sigma'} \sum_{m\neq m'}(U{-}2J{-}J\delta_{\sigma,\sigma'})\hat{n}_{im\sigma} \hat{n}_{im' \sigma'}  \\ \nonumber
&-J\sum_{ i m\neq m'}(c^{\dagger}_{im\uparrow} c^{\dagger}_{im\downarrow} c_{im'\uparrow} c_{im'\downarrow}{+}c^{\dagger}_{im\uparrow} c_{im\downarrow} c^{\dagger}_{im'\downarrow} c_{im'\uparrow}). 
\end{align} 
Here $-t^{i,i'}_{mm'}$ ($i\ne i'$) are the LDA hopping integrals \cite{sonote} from orbital $m$ on site $i$ to orbital $m'$ on  site $i'$. 
They are obtained in a localized Wannier function basis 
via the linearized augmented plane wave method \cite{BLAHA1990399,PhysRevB.56.12847,wannier90}.
The operator $c^{\dagger}_{im\sigma}$ ($c_{im\sigma}$) creates (annihilates) an electron with spin $\sigma$ in Wannier state $m$ at site $i$, and $n_{im\sigma}=c^{\dagger}_{im\sigma}c_{im\sigma}$.  For the screened Coulomb parameters 
we use values established in previous works \cite{PhysRevB54.5368,3d1a,3d1b,Molly2007}:  $U{=}5$ eV with $J{=}0.68$ eV for YVO$_3$ and LaVO$_3$, 
and $J{=}0.64$ eV for YTiO$_3$ and LaTiO$_3$. We then solve this model using dynamical mean-field theory (DMFT). 
The quantum impurity solver  is the generalized hybridization-expansion continuous-time quantum Monte Carlo method \cite{RevModPhys.83.349}, in the implementation of Refs.~\cite{lamno3c,julian1,julian2}. 

\begin{figure}
\centering  %
\includegraphics[width=0.49\textwidth]{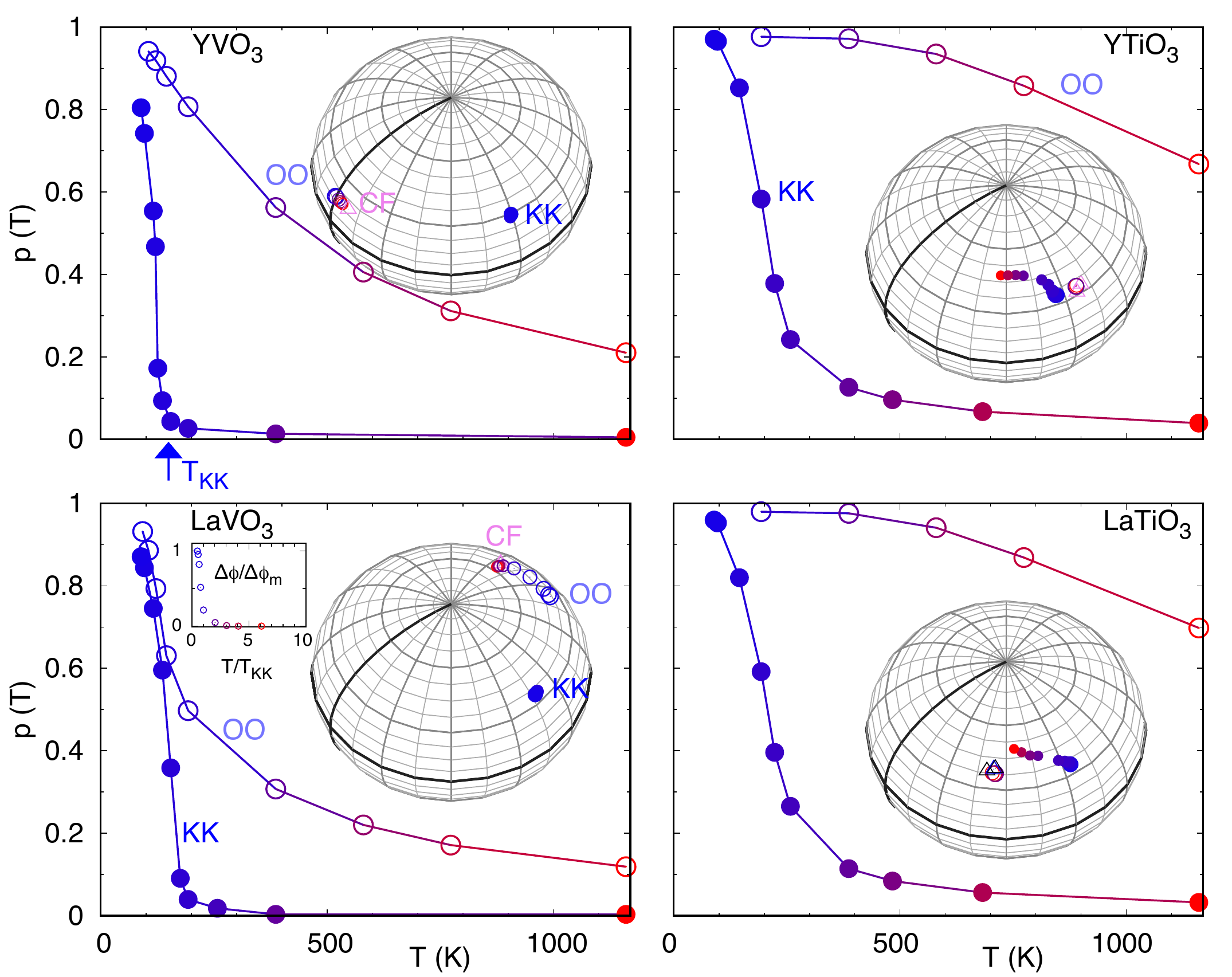}
\caption{\label{dmftt2g}
Lines: Orbital polarization $p(T)$, defined as $p(T){=}(-1)^n({(n_1{+}{n}_2)}/{2}{-}n_3)$, as a function of temperature $T$; here $n_i$
is the occupation of natural orbitals, with $n_{i+1}{<}n_i$  for $n{=}2$ and  $n_{i+1}{>}n_i$  for $n{=}1$. 
Bloch spheres: angles defining the least $(n{=}2)$ or most ($n{=}1$) occupied orbital,
 $|\theta,\phi\rangle{=}\sin\theta \cos\phi |xz\rangle {+} \cos\theta |xy\rangle {+} \sin\theta \sin \phi |yz\rangle$.  
Empty circles (OO): experimental structure. Filled circles (KK): Idealized case with no CF splitting.
Triangles (CF):  crystal-field orbital. 
Left: vanadates. Right: titanates. %
Equatorial line: $\theta{=}90^\circ$.  Thick-line meridian: $\phi{=}0^\circ$.
The equivalent solution at $|\pi{-}\theta,\phi{+}\pi\rangle$  is on the far side of the sphere.
{For LaVO$_3$, an inset shows explicitly the sudden variation of  $\phi$ across $T_{\rm KK}$;
$\Delta \phi=\phi{-}\phi_{\rm CF}$ is the difference with respect to the CF value of $\phi$.}}
\end{figure}

The main results  are summarized in  Fig.~\ref{dmftt2g}.  
We start from  the experimental structure, with full CF splitting (empty circles). 
For the titanates, the orbital polarization
$p_{\rm OO}(T)$  is already very large at 1000 K;
instead, in  the vanadates, and   LaVO$_3$ in particular,
$p_{\rm OO}(T)\longrightarrow 1$ at much lower temperatures  \cite{Molly2007}.
The OO ground state, for the $t_{2g}^2$ case, is characterized, a  given site, by the least occupied (or hole) natural orbital 
\begin{align}\label{state}
|\theta,\phi\rangle %
=&\sin\theta \cos\phi |xz\rangle + \cos\theta |xy\rangle + \sin\theta \sin \phi |yz\rangle,
\end{align}
represented via open circles on the Bloch spheres in the figure.
The hole orbitals at the neighboring sites, yielding the spatial arrangement of orbitals, can be obtained from (\ref{state}) using symmetries \cite{suppl}.
For $t_{2g}^1$ systems (\ref{state}) represents  the most occupied orbital. As the figure shows, in the $t_{2g}^1$ case, 
at any temperature $|\theta,\phi\rangle_{\rm OO}$ is very close to $|\theta,\phi\rangle_{\rm CF}$, the lowest energy
crystal-field orbital (open triangle). 
Switching to $t_{2g}^2$ systems,
for  YVO$_3$, 
the hole orbital  is $|\theta,\phi\rangle_{\rm{OO}} \sim |72^\circ,-1^\circ \rangle$, %
quite close to the crystal-field state with the highest energy,
$|\theta,\phi\rangle_{\rm{CF}} \sim |71^\circ,9^\circ\rangle$.
Up to here, the results  conform to the established picture: the CF splitting, enhanced by Coulomb repulsion, determines
the shape of the ordered state
\cite{prl2008,prl2010,autieri,lamno3c,tote,julian1}.

The conclusion changes dramatically as soon as we turn to  LaVO$_3$. 
Here, $p_{\rm OO}(T)$ has a sharp turn upwards at $T_{\rm KK}{\sim} T_{\rm OO}{\sim}190$~K.
Furthermore, lowering the temperature, the hole orbital $|\theta,\phi\rangle_{\rm{OO}}$ turns markedly away from
the CF high-energy state, 
$|\theta,\phi\rangle_{\rm{CF}}\sim|142^\circ,25^\circ\rangle
$, showing that the ordering mechanism works against the crystal-field splitting.
At the lowest temperature we could reach numerically, $|\theta,\phi\rangle_{\rm{OO}} \sim|130^\circ,-8^\circ \rangle$.
This can be seen %
on the Bloch sphere, where the empty circles  move away from the triangle {as well as in the inset showing the sudden variation of  the most occupied orbital across $T_{\rm KK}$}. 

Next we analyze the results in the zero CF splitting limit, which yields
the pure Kugel-Khomskii transition.  The results are shown as filled circles in  Fig.~\ref{dmftt2g},
and the orbital polarization curve is $p_{\rm KK}(T)$.
For the titanates,  $p_{\rm KK}(T)$ exhibits a small 
tail at high temperature and then sharply rises at  $T_{\rm{KK}}{\sim}300$~K;
at this transition $p_{\rm OO}(T)$ has, however, long saturated.
For the vanadates
the rise in  $p_{\rm KK}(T)$ is much sharper,
and at a markedly  lower temperature,  $T_{\rm{KK}}{\sim}150$~K in YVO$_3$ and 
 $T_{\rm{KK}}{\sim}190$~K in LaVO$_3$, while the high-temperature tail is virtually absent. Furthermore, 
for LaVO$_3$,  the figure shows that $p_{\rm OO}(T){\sim} p_{\rm KK}(T)$  for  temperatures sufficiently below $T_{\rm KK}{\sim}T_{\rm OO}$.
At the same time, decreasing the temperature,
the  OO hole orbital for the experimental structure (empty circles), 
rapidly moves towards the KK hole  (filled circles).
These results together identify LaVO$_3$ as the best known representation of a Kugel-Khomskii system, i.e., a system {hosting an orbital-ordering 
KK-driven phase transition. The KK super-exchange interaction both determines the value of $T_{\rm KK}{\sim}T_{\rm OO}$
and pulls the hole away from the crystal-field orbital  towards the KK orbital.  }
YVO$_3$ is on the borderline, but still on the side where the Coulomb-enhanced CF interaction dominates; that means, 
the hole stays close to the crystal field orbital even below $T_{\rm  OO}$.  Furthermore, for YVO$_3$, the critical temperature $T_{\rm KK}\sim 150$~K itself
is smaller that the $T_{\rm str}\sim 200$~K, the temperature at which the orthorhombic-to-monoclinic phase transition occurs. 
For LaVO$_3$ the opposite is true $(T_{\rm KK}{\sim}T_{\rm OO} {>}T_{\rm str})$.
\begin{figure}[t]
\centering  
\rotatebox{270}{\includegraphics[width=0.7\textwidth]{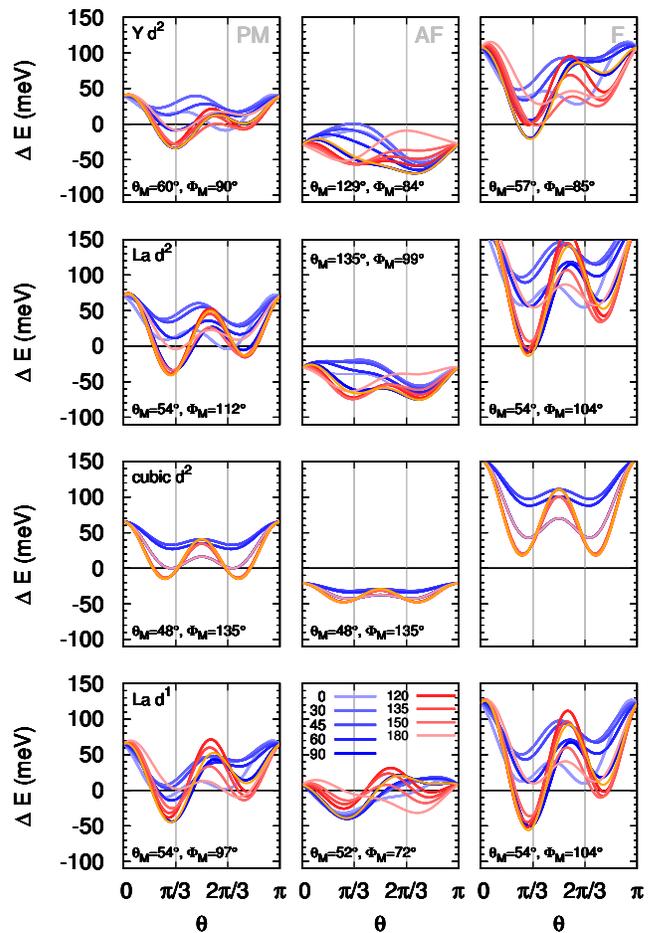}}\\
\caption{ \label{ene} 
Total super-exchange energy gain for YVO$_3$ (top panel) and LaVO$_3$ (second panel from the top)
in the para-(PM), antiferro- (AF) and ferro-magnetic (F)
case, GdFeO$_3$-type structure. Lines: different $\phi$ values in the $(0,\pi)$ interval, see caption. Orange curve: $\phi_{\rm M}$, yielding the minimum.
Third panel from the top: cubic case.
Bottom panel: energy gain for the hopping integrals of LaVO$_3$, but in a hypothetical $t_{2g}^1$
 configuration.  %
}
\end{figure}
\begin{figure}[t]
\centering  
\includegraphics[width=\columnwidth]{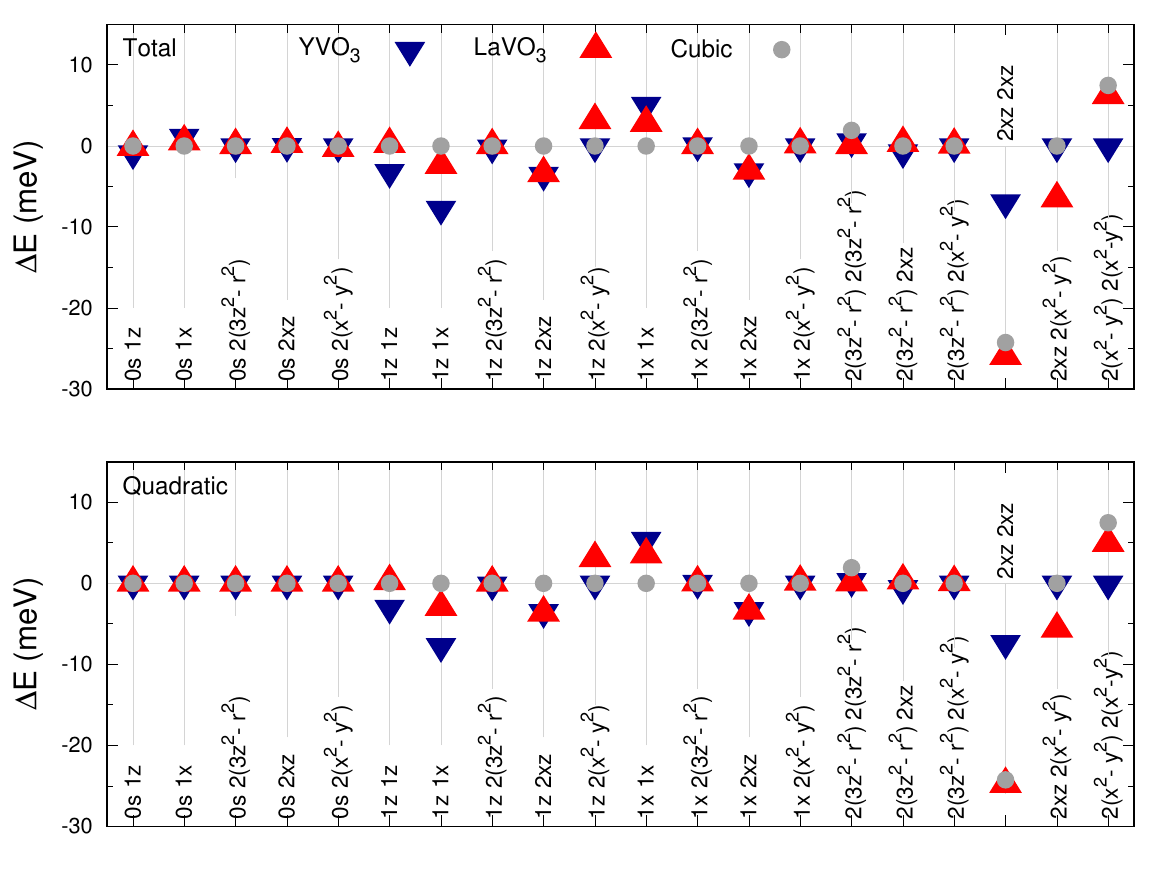}
\vspace*{-5ex}
\caption{ \label{DEt2g2} 
Decomposition of $\Delta E (\theta,\phi)$
at the angles $(\theta_M,\phi_M)$ that yield the absolute minimum in the paramagnetic case.
Top: result from the total $\Delta E (\theta,\phi)$. Bottom:  only quadratic terms are included.
The components are specified on the horizontal axis;
 missing non-diagonal terms  can be obtained via the transformation $r\mu\leftrightarrow r'\mu'$. }
\end{figure}

\section{Super-exchange  Hamiltonian and energy surfaces}
Further support for these conclusions comes from the analysis of super-exchange energy-gain surfaces,
Fig.~\ref{ene}.
We obtain them adopting the approach
recently introduced in Refs.~\cite{xj2020,tobepublished}, representing
 the multi-orbital KK super-exchange Hamiltonian via its irreducible-tensor decomposition
\begin{align} \label{SE}
\hat{H}_{\rm SE}
=\frac{1}{2}\sum_{i,j}\sum_{\mu\mu'} \sum_{r,r'}\hat{\tau}_i^{r,\mu; q\nu} D^{ij ; q\nu}_{r\mu,r'\mu'} \hat{\tau}_{j}^{r',\mu'; q\nu}\;
\delta_{q,0}\delta_{\nu,0}.
\end{align}
Here $r=0,1,2$ is the orbital rank and $\mu$ the associated components;
the spin rank is $q$ with components $\nu$.
The analytic expressions of the tensor elements 
can be found in  Refs.~\cite{xj2020,tobepublished}.
The terms with $q{=}q'{=}0$ and $r{=}0,r'{\ne}0$ (or viceversa) describe the (linear) orbital Zeeman interaction \cite{julian1}.
These contributions behave as a {site-dependent} crystal-field splitting, since the $r{=}q{=}0$
tensor operator only counts the number of electrons on the neighboring site,
here $n{=}2$. 
The  $r{\ne}0,r'{\ne}0$ quadratic terms are those that can actually lead to a phase transition.
Terms with $q{=}0$ are purely paramagnetic, those with $r{=}0$ purely paraorbital.
In Fig.~\ref{ene} we show results for the {\em ab-initio} hopping integrals  and, for reference, an idealized cubic case. 
Comparing the results for the PM phase with those
in Fig.~\ref{dmftt2g},  one may see that the angles $\theta_{\rm{M}}, \phi_{\rm{M}}$ minimizing  $\Delta E(\theta,\phi)$
yield  $|\theta,\phi\rangle_{\rm KK}$, the state obtained in LDA+DMFT calculations
 in the zero CF limit.
The maximum quadratic SE energy gain, $|\Delta E(\theta_M,\phi_M)|$, increases from
 $\sim{35}$~meV for YVO$_3$ to $\sim{41}$~meV for LaVO$_3$,
to $\sim{46}$ meV for LaTiO$_3$ and YTiO$_3$,  explaining
the progressive increase in $T_{\rm KK}$ obtained in the DMFT calculations.

Fig.~\ref{DEt2g2} shows the  various  contributions to $\Delta E(\theta_{\rm{M}}, \phi_{\rm{M}})$.
The most important are the quadratic SE terms, those which can lead to a transition.
For YVO$_3$ several channels have comparable weight -- similarly to the titanates \cite{suppl}.
In contrast, for LaVO$_3$,  a single term,
the $\hat{\tau}^{2,xz}_i\hat{\tau}^{2,xz}_j$ interaction, dominates, as in the cubic limit;
 the total energy gain from quadratic SE terms
is, however, significantly larger than in the cubic limit, since small positive and negative contributions
cancel out.

The linear orbital Zeeman terms \cite{suppl},
while  not giving rise to a phase transition,  
can generate 
a finite polarization tail already for $T>T_{\rm KK}$ \cite{julian1}, which can either assist or hinder the transition.
For the titanates  there is a sizable tail in the $p_{\rm KK}(T)$ curves in  Fig.~\ref{dmftt2g};
furthermore, the linear terms cooperate with the quadratic terms, favoring the occupation of the same orbital.
For the vanadates, instead, the tail  is negligible. 
The suppression of the orbital Zeeman effect turns out to be due to  filling, rather than to the band structure:
Assuming identical hopping integrals in the two families of compounds, $D^{ij \, 00}_{00,r\mu}(n{=}1)/ D^{ij \, 00}_{00,r\mu}(n{=}2){=}{\cal W}_1/{\cal V}_1$, as defined in \cite{tobepublished}, Table III;
the right hand side  depends only on $J/U$.
The  
 prefactor ${\cal W}_1$  for the $t_{2g}^1$ configuration can be sizably larger
than the corresponding ${\cal V}_1$ for the $t_{2g}^2$ case -- up to a factor of 10 for realistic $J/U$ values \cite{tobepublished}.
This can be seen comparing the La $d^2$ and $d^1$ panels in Fig.~\ref{dmftt2g}.

Summarizing, in the PM phase,  quadratic SE interactions are weaker in the vanadates than in the titanates, and the orbital Zeeman
terms are negligible.  LaVO$_3$ is, however, characterized by a very small CF splitting;
the KK Hamiltonian form is close  to the cubic limit, but the distortions actually increase the energy gained 
from ordering the orbitals. This makes  LaVO$_3$ unique,
and results in  low-temperature orbital physics being controlled by
super-exchange interactions. 

\section{Orbital ordering and the G-type anti-ferromagnetic phase} 
\begin{figure}[t]
\centering  
\rotatebox{0}{\includegraphics[width=0.45\textwidth]{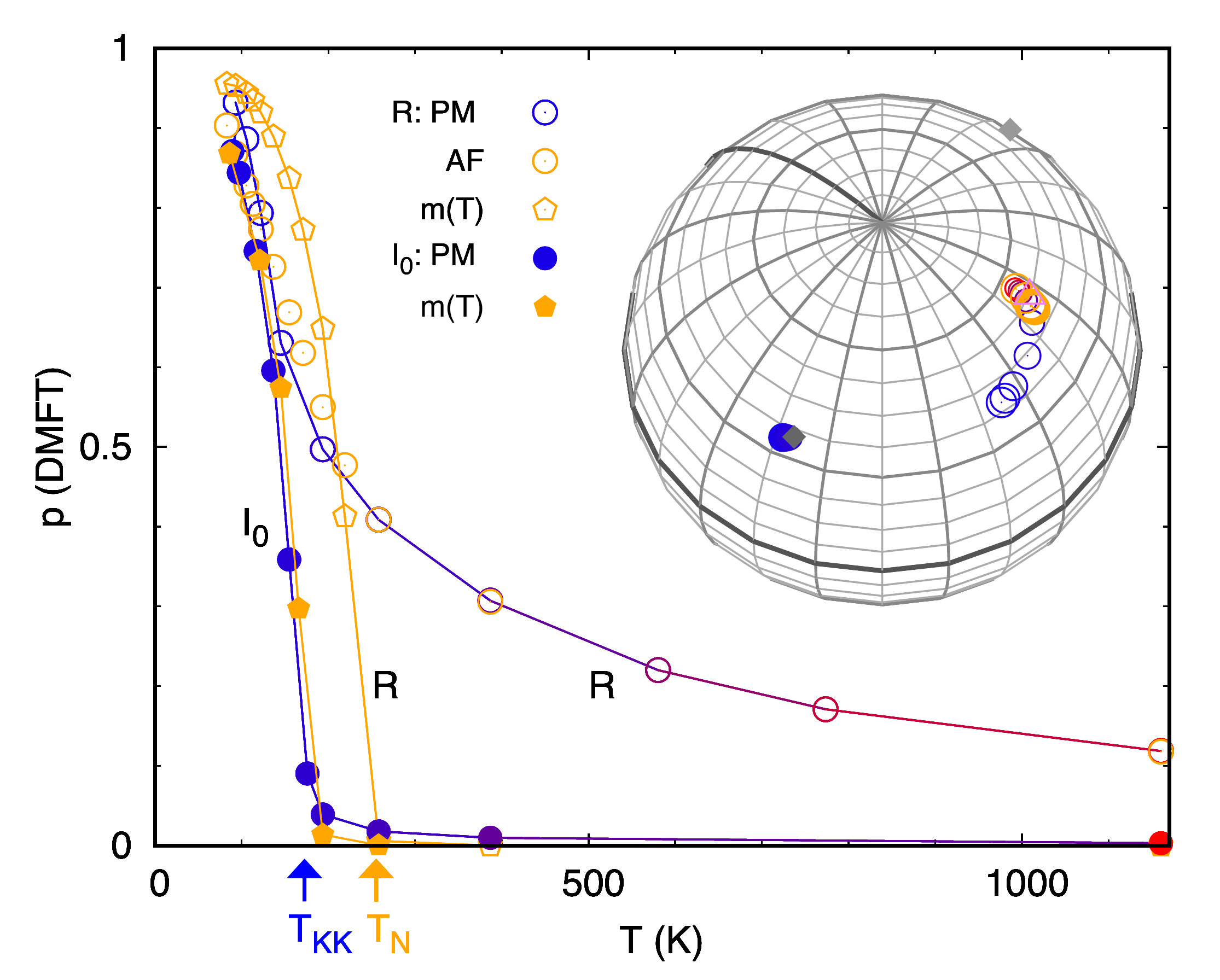}}\\ 
\caption{ \label{dmftafm} LaVO$_3$, orbital and magnetic
transition. 
Circles: orbital polarization. Pentagons: magnetization.
Empty symbols: DMFT for experimental structure ($R$). 
Full symbols: DMFT for an ideal structure without CF splitting  ($I_0$).  
%
%
Triangle: LDA. 
Rhombi: angles yielding the maximum SE energy-gain (AF case),
obtained from quadratic terms (dark grey) and  linear terms (light grey) only.
}
\end{figure}

To obtain the final confirmation of  the dominant role of super-exchange in  LaVO$_3$ 
we perform calculations allowing for G-type antiferromagnetism.
For conventional orbitally ordered materials, where OO is driven by the Coulomb-enhanced crystal field, 
$T_N\ll T_{\rm OO}$. In   LaVO$_3$, instead,  we find that  $T_N$ and $ T_{\rm OO}$ are comparable.
In fact, 
as illustrated  in Fig.~\ref{dmftafm}, the magnetic ordering happens already
{\em above} the orbital ordering transition,
$T_{N}>T_{\rm OO}\sim T_{\rm KK}$. 
To explain this remarkable result, we return to the SE energy surfaces
in Fig.~\ref{ene}, 
and compare the PM case,
left column, to  the AF case, center column. 
The figure shows that the AF curves are shifted uniformly downwards. 
This is due to the paraorbital   ($r{=}0$) term with spin rank $q{=}1$; comparing to the bottom row of the figure,
one may see  
that the latter is much larger in the  $t_{2g}^2$ than in the $t_{2g}^1$ configuration.
Furthermore, the quadratic energy gain for orbital ordering alone (obtained subtracting the paraorbital term) decreases going from the PM to the AF case; at the same time, the orbital Zeeman linear terms increase in importance,
but favor   $ \phi_{M'}\sim \phi_M{+}180^\circ$, hence competing with the quadratic terms.  
Thus,   at a given finite temperature, with respect to the PM case, the OO hole orbital (orange open circle on the sphere) remains closer to the highest-energy CF orbital (triangle).
Importantly, we obtain such a behavior only for LaVO$_3$; in the case of YVO$_3$, for the experimental structure we do not find any magnetic phase above $T_{\rm str}$ or $T_{\rm KK}$,
 in line with experiments.

\section{Conclusion}
In conclusion, we have  identified orthorhombic LaVO$_3$ as a rare case  {of a system hosting a Kugel-Khomskii  orbital-ordering  transition.} %
The relevance of SE in $t_{2g}^2$ vanadates has been  previously suggested based on idealized model
calculations \cite{Khal},  however, within a  strong $xz/yz$ fluctuations picture. 
In fact, we have shown that  orbital fluctuations, large at room temperature, are suppressed when approaching $T_{\rm str}$. 
Furthermore, we have shown that
SE is key only for LaVO$_3$. In all other cases considered,
the conventional picture of the Coulomb-enhanced CF splitting applies.
This is further confirmed by the fact that $T_N\sim T_{\rm OO}$ only in orthorhombic LaVO$_3$.
More strikingly, we find that
$T_N> T_{\rm KK}\sim T_{\rm OO}$, opposite to what 
happens in conventional OO materials. 
Taking into account that DMFT, as all mean-field theories,  somewhat overestimates  ordering temperatures, our results provide a natural explanation for the proposed picture of orbital order and weak G-type magnetism (or short range spin-orbital  
order) right above the structural orthorhombic-to-monoclinic  transition \cite{Ren2003,Zhou2008,Tung2008}.  

\acknowledgments
We would like to acknowledge computational time on JURECA and the RWTH-Aachen cluster
via JARA, which was used for the actual computations, as well as computational time on JUWELS, which was used in particular for code development.


\begin{thebibliography}{99}
\bibitem{sonote} The small spin-orbit coupling and related entanglement effects are neglected.
\end{thebibliography}
\end{document}